\documentclass[aps,prb,showpacs,
floatfix,
amsmath,amssymb,
               reprint]{revtex4-1}
\usepackage{bbold}
\usepackage{color}
\usepackage{graphicx}
\usepackage{natbib}
\usepackage{setspace}
\usepackage{amsmath}
\usepackage{amssymb}
\usepackage{verbatim}
\usepackage{bibentry}
\usepackage{physics}
\usepackage{hyperref}
\usepackage{cancel}
\def\etal{{\em{et al. }}}

\def\nnu{{\nonumber}}
\def\bk{{\mathbf{k}}}

\definecolor{scarred}{rgb}{0.75,0.0,0.0}

\begin{document}


\title{Steady-state dc transport through an Anderson impurity coupled to leads with spin-orbit coupling}
\author{Anirudha Mirmira}
\affiliation{Jawaharlal Nehru Centre for Advanced Scientific Research, Bangalore 560 064, India.}
\author{N.\ S.\ Vidhyadhiraja} \email{raja@jncasr.ac.in}
\affiliation{Jawaharlal Nehru Centre for Advanced Scientific Research, Bangalore 560 064, India.}
\begin{abstract}
We study the steady-state dc transport characteristics of a system comprised of an interacting quantum dot, modeled as an Anderson impurity, coupled to two metallic non-interacting leads with Rashba spin-orbit coupling (SOC), using an interpolative perturbative approach (IPA). The single-particle spectra, current, and differential conductance are obtained in weak and strong coupling regimes over a wide range of SOC and bias values. Extensive benchmarking of the IPA validates the method
in the linear as well as non-linear response regime.
The  universal, zero bias ($V_{sd}=0$) peak with a width proportional to the Kondo scale ($T_K$) and two non-universal finite bias peaks around $V_{sd}=\pm U$ in the zero temperature differential conductance show a clear separation with increasing $U$ or increasing SOC.
In the strong coupling regime, increasing temperature induces melting of the zero bias peak, leading to a crossover from a three-peak conductance to a two-peak conductance. Recent experiments find the emergence of a two-peak structure by increasing SOC at a fixed temperature. Our results appear to provide a qualitative explanation of these observations as a SOC tuned crossover from weak/intermediate to strong coupling, and a simultaneous crossover from low $T/T_K$ to high $T/T_K$ ratio. We also reproduce the experimentally observed temperature dependence of the zero bias conductance.  
\end{abstract}
\maketitle

\section{Introduction}

    The Kondo effect and its interplay with bias in the leads is a rich and well-studied problem. This non-equilibrium Kondo effect has been observed in a wide variety of systems such as quantum dots\cite{goldhaber1998kondo,cronenwett1998tunable,chorley2012tunable}, molecular systems\cite{liang2002kondo,yu2004kondo}, carbon nano tubes\cite{grigorian1998giant,nygaard2000kondo,eichler2009tuning,laird2015}, and quantum point contacts\cite{van1988quantized,thomas1996possible,brun2014wigner,heedt2016ballistic,smith2022electrically}. Correspondingly, experiments also observe a wide range of phenomena in these systems such as quantum interference\cite{aradhya2013single}, spin-selective trasport\cite{kobayashi2010kondo}, etc. The single impurity Anderson model (SIAM) has been the standard paradigm for modeling such systems both in and out of equilibrium. The SIAM in equilibrium has been studied extensively using methods such as the Bethe ansatz\cite{tsvelick1983exact} and numerical renormalization group\cite{krishna1980renormalization,ralfbullanrg}.
    
    In recent years, a whole range of theoretical methods has been developed and used to study Kondo systems out of equilibrium. However, unlike in the case of equilibrium, where a reasonable consensus has been achieved, the physics of nonequilibrium interacting systems is still under debate. In the case of the SIAM, the splitting of the Kondo peaks, the effects of magnetic field and spin-orbit coupling (SOC) on the Kondo effect, and the effects of bias and temperature have been studied using various methods such as numerical renormalization group\cite{anders2008steady,anders2005real,fritsch2010nonequilibrium,minarelli2022linear}, non-crossing approximation\cite{wingreen1994anderson}, perturbative methods\cite{fujii2003perturbative,lopes2019spin,joura2015long}, quantum Monte Carlo methods\cite{werner2010weak}, exact diagonalization methods\cite{nuss2013steady}, Bethe ansatz\cite{mehta2006nonequilibrium}, hierarchical equation of motion\cite{cheng2015time} and master equation approaches\cite{dorda2014auxiliary} among many others. 
    
    SOC is an important parameter, and is well known to be an integral aspect of exotic materials such as topological insulators, and topological magnets and is useful in many spintronics applications. Quantum dot systems have been studied both with SOC on the dot and with SOC in the leads. Experiments with the SOC on the dot include directly measuring the SOC in two-atom quantum dots\cite{fasth2007direct}, the observation of an anomalous Josephson current\cite{zazunov2009anomalous}, and controlling the SOC using magnetic field direction \cite{tanttu2019controlling} among many others. The case with SOC in the leads has also been studied with phenomena such as the gate voltage tunability of SOC\cite{guo2021zero,smith2022electrically}, magnetotransport\cite{schapers2004effect}, etc. having been observed. 
   
    In equilibrium, the effect of the SOC on the Kondo resonance has been theoretically studied in detail\cite{sandler,wong2016influence,bonvca2011kondo}. A driven quantum dot, coupled to normal metallic leads, with SOC on the dot has been investigated using the finite-$U$ slave boson method\cite{lue2007kondo} and the findings show that the Rashba spin-orbit coupling (RSOC) introduces new conductance peaks next to the Kondo peak while suppressing the Kondo peak. Quantum wire systems have also been studied and interesting observations, such as the destruction of spin accumulation due to an impurity\cite{zhang2011rashba} and multichannel  
    effects\cite{gelabert2010multichannel} have been made. A quantum dot connected to a nanoribbon with SOC has also been studied using the Hubbard III approximation\cite{lopes2019spin}.

    A powerful tool in experimentally studying the effect of bias on quantum systems is the paradigm of quantum point contacts (QPCs). QPCs are realized by constricting a two dimensional (2D) electron gas between contacts which have a source-drain and gate voltage applied. Multiple experiments on QPCs have observed quantized conductance along with some well-known anomalies \cite{topinka2000imaging,meir2002kondo,iqbal2013odd}. Recent experiments in systems with  quantum point contacts have also seen the interplay of SOC and Kondo physics. In particular, Smith \etal\cite{smith2022electrically}, have realized a QPC setup where they can tune the SOC and have observed a two-peak conductance in the large SOC regime along with an increase in the zero bias conductance (ZBC) with temperature.

    In this paper, we study the interplay between Rashba spin-orbit coupling on the leads, a constant bias and interactions on the steady-state dc transport through a quantum dot system. As mentioned before, the effect of Rashba SOC in the leads on the Kondo effect in equilibrium has received significant attention\cite{sandler,wong2016influence,bonvca2011kondo}. When subjected to a dc bias, the studied system will be out of equilibrium and the interplay of bias and SOC in the leads becomes an important consideration. We employ the interpolative approximation(IPA)\cite{aligiamain} , which reduces in the particle-hole (p-h) symmetric limit to the Keldysh second-order perturbative theory (KPT2), to investigate transport through the dot with Rashba SOC on normal metallic two-dimensional leads. Since the method is approximate, we begin with a benchmarking of the method against exact methods such as the time-evolving block decimation (TEBD)\cite{nuss2013steady} and identify the regimes within which the results from KPT2 are reliable. We have investigated the effect of SOC on (i) universality and scaling in the linear response regime and, subsequently, on (ii) differential conductance in the non-linear bias regime. We find the equilibrium universal scale, i.e. the quasiparticle weight controls the extent of the linear response regime and is also crucial in determining the thermal scaling of the system when comparing the theoretical and experimental results. The features in differential conductance are investigated in a wide parameter space, and characteristic signatures of the interplay of bias, interactions, Rashba SOC and temperature are identified. Finally, we compare some of our results corresponding to the strong coupling regime to recent experiments, and offer a qualitative explanation for some of the observations.

We begin with a brief outline of the model and formalism in the next section. Subsequently, we present our results in three parts  in Sec. III. A short discussion and the conclusions are presented in the final section. A few details regarding the derivation of some results are presented in an appendix.

\section{Formalism}

The Hamiltonian for a quantum dot system connected to two leads with Rashba spin-orbit coupling (RSOC) can be written as 
\begin{equation}
    H = H_{0}+H_{\rm d}+H_{\rm RSOC} + H_{\rm hyb}\,,
\end{equation}
where the two-dimensional conduction band reservoirs (L/R) are represented by
$H_0 = \sum_{\alpha\bk\sigma}\epsilon_{\alpha\bk}^{\phantom{\dagger}} c^\dagger_{\alpha\bk\sigma} c_{\alpha\bk\sigma}^{\phantom{\dagger}}$ and the Hamiltonian for the quantum dot is given  by
$H_{\rm d} = \sum_{\sigma}\epsilon_{d} d^{\dagger}_{\sigma}d_{\sigma}^{\phantom{\dagger}}+Un_{d\uparrow}n_{d\downarrow}$,
where $\alpha=L/R$ and $\sigma=\uparrow/\downarrow$ are the lead and spin indices respectively. 
The RSOC term is represented by \cite{sandler}
\begin{equation}
 H_{\rm RSOC}=\sum_{\alpha\bk}\lambda \psi^{\dagger}_{\alpha\bk}(\bk\times \vec{\sigma})_z\psi_{\alpha\bk}\,,
\end{equation}
where $\bk=(k_x,k_y)$, and $\psi^{\dagger}_{\alpha\bk}=(c^{\dagger}_{\alpha\bk,\uparrow}\;c^{\dagger}_{\alpha\bk,\downarrow})$.
Finally, the hybridization between the quantum dot and the leads is given by
\begin{equation}
    H_{\rm hyb}=\sum_{\alpha\bk\sigma} \left( V_\bk c^\dag_{\alpha\bk\sigma}d_\sigma^{\phantom{\dagger}} + {\rm h.c}\right)\,.
\end{equation}
The conduction band terms, namely, $H_0$ and $H_{\rm RSOC}$, may be combined \cite{sandler}, which leads to the emergence of chiral conduction bands. This is accomplished using an angular momentum expansion for the conduction band operators as
\begin{equation}
\label{eq:AMtrans}
    c_{\bk\sigma} = c_{k_x k_y\sigma} = \frac{1}{\sqrt{2\pi k}}\sum_{m=-\infty}^{\infty}
    c_{km\sigma}\exp(i m \theta_\bk)\,,
\end{equation}
where $k=|\bk|$, with inverse transform being defined as $c_{km\sigma}=\sqrt{k/2\pi}\int^{2\pi}_{0} d\theta_{\bk} c_{\bk \sigma} e^{-im\theta_{\bk}}$.

Substituting the above expansion (Eq. ~\ref{eq:AMtrans}) into the Hamiltonian, and assuming an isotropic dispersion, given by $\epsilon_\bk = \hbar^2 k^2/2m$, the following form of the Hamiltonian is obtained\cite{sandler}:
\begin{equation}
   \begin{split}
    H &=\sum_{khm} \tilde{\epsilon}_{k h} (c_{kh}^{m+\frac{1}{2}})^\dagger c_{kh}^{m+\frac{1}{2}} \\
    & + \sum_{khm} \delta_{m,0} \tilde{V_k}\left((c_{kh}^{m+\frac{1}{2}})^\dagger d_\uparrow + h(c_{kh}^{m-\frac{1}{2}})^\dagger d_\downarrow + {\rm h.c}\right) \\
    &+ H_{d}\,,
    \end{split}
\end{equation}
where $h=\pm 1$ is an emergent chiral quantum number. This also allows us to define an emergent angular momentum quantum number, given by $j_m=m+\frac{h}{2}$, with only the bands corresponding to $j_m=\pm\frac{1}{2}$ coupling to the dot, while the rest of the bands are decoupled.
The renormalized dispersion $\Tilde{\epsilon}_{kh}$, now depends on $h$ and
the spin-orbit interaction $\lambda$, as given by the expression $\tilde{\epsilon}_{kh} = (\epsilon_{k}+ h\lambda k)/k 
=\tilde{\epsilon}_{k} + h\lambda$. 
For a free-electron like dispersion, $\tilde{\epsilon}_{k}$ will be linear in $k$. Hence RSOC introduces a Zeeman-type splitting of the conduction band, without breaking the time-reversal symmetry. 
Further, the hybridization matrix elements are assumed to be isotropic, i.e, $V_\bk = V_k$, and $\tilde{V}_k = V_k\sqrt{2\pi/k}$.

\begin{figure}
    \includegraphics[width=\linewidth,trim=0 0 0 0, clip]{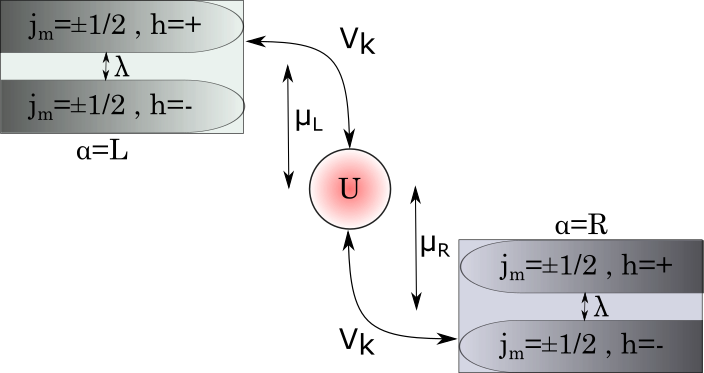}
     \caption{Schematic of the correlated quantum dot connected to two leads, split further by spin-orbit coupling into chiral bands, subject to a voltage bias, $V_{sd}=\mu_L - \mu_R$.}
    \label{fig:schem_sys}
\end{figure}

In order to study the interplay between the RSOC and a constant voltage bias, we consider the left and right leads to have a lead-dependent chemical potential $\mu_\alpha$ applied to all the emergent $h,j_m$ channels in each lead, such that the voltage bias is given by $V_{sd}=\mu_L-\mu_R$. This leads to the Hamiltonian
\begin{equation}
        H_{eff} = \sum_{\alpha=L,R} H_\alpha +H_{hyb}+ H_{dot}\,,
\end{equation}
where the individual terms are given by
\begin{align}
    \label{eq:SOanderson}  
       H_{\alpha}& = \sum_{khj_m}\tilde{\epsilon}^{\phantom{\dagger}}_{k \alpha h} c^\dagger_{\alpha khj_m}c^{\phantom{\dagger}}_{\alpha khj_m}\,, \\
        H_{hyb}&= \sum_{\alpha kh}\tilde{V}^{\phantom{\dagger}}_{k\alpha}\big[c^\dagger_{\alpha kh\frac{+1}{2}}d^{\phantom{\dagger}}_\uparrow +hc^\dagger_{\alpha kh\frac{-1}{2}}d^{\phantom{\dagger}}_\downarrow +h.c.\big]\,, \\
H_{dot} &= \sum_{\sigma}\epsilon_d n_\sigma + Un_\uparrow n_\downarrow\,,
\end{align}
with $\tilde{\epsilon}_{k\alpha h}=\tilde{\epsilon}_{kh}+\mu_\alpha$. The model described above can be visualized by the schematic in Fig. \ref{fig:schem_sys}. We note that this model was investigated using a quantum master equation (QME) approach in a recent work~\cite{smith2022electrically}, and the results were used to explain specific experimental observations of differential conductance reported in the same work. The present work, using the IPA, provides a different perspective and fresh insight into these experiments~\cite{smith2022electrically} and hence may be viewed as complementary to the QME results.

In the presence of a finite bias ($V_{sd}\neq 0$), the steady
state current ($J$) can be computed using the symmetrized Meir-Wingreen formula,
given by \cite{meirwingreen}
\begin{align}
\label{eqn:curr}
    J = & \frac{2i e}{h}\sum_\sigma\int d\omega\,\Big[ \left(\Gamma_L(\omega)-\Gamma_R(\omega)\right)G^<_{d\sigma}(\omega) \nnu \\
    + & 2i \left(f_L(\omega)\Gamma_L(\omega) - f_R(\omega)\Gamma_R(\omega)\right){\rm Im}G^r_{d\sigma}(\omega)\Big]\,.
\end{align}
In the above, the $G^<_{d\sigma}$ and $G^r_{d\sigma}$ represent the lesser and retarded dot Green's functions computed within Keldysh formalism, and $f_\alpha$ is the Fermi-Dirac distribution function of the $\alpha^{\rm th}$ lead. The $\Gamma_\alpha(\omega)$ represents the dot-lead coupling
and is given by
\begin{equation}
\Gamma_\alpha(\omega) = -{\rm Im} \Delta_\alpha(\omega)\,,
\end{equation}
where the hybridization function for the $\alpha^{\rm th}$ lead, $\Delta_\alpha(\omega)$, is given by
\begin{equation}
\begin{split}
\Delta_\alpha(\omega) & = \sum_{kh}\frac{{\tilde{V}_k}^2}{\omega^+ -\epsilon_{k\alpha h}} \\
            & = \sum_k {\tilde{V}_k}^2 \left[\frac{1}{\omega^+ - \tilde{\epsilon}_{k\alpha} + \lambda}
            + \frac{1}{\omega^+ - \tilde{\epsilon}_{k\alpha} - \lambda}
            \right]\\
            & = \sum_h \Delta_{h\alpha}(\omega)\,=E_\alpha(\omega)-i\Gamma_\alpha(\omega).
\end{split}
\label{eq:hybksum}
\end{equation}

With a suitable choice of the $\tilde{V}_k$-dependence on $k$ as $\tilde{V}_k^2 = V_0^2f(k)$, we can transform the summation over $k$ into an integral that has the form of a Hilbert transform, and hence the hybridization function can be obtained as:
\begin{equation}
    \Delta_{h\alpha}(\omega) = V_0^2 {\rm H}\left[\omega^+-\mu_\alpha-h\lambda\right]
    \label{eq:dh}
\end{equation}
where ${\rm H}[z]$ is the Hilbert transform with respect to a "density of states" (DOS), $\rho_0(\epsilon)$, given by
\begin{equation}
    {\rm H}[z]=\int\,d\epsilon\,\frac{\rho_0(\epsilon)}{z-\epsilon}\,.
\end{equation}
Note that the $\rho_0(\epsilon)$ in the above equation stems from the $k$-dependence of the hybridization matrix element. The conduction band density of states is already incorporated through the electron dispersion of a free-electron form (see the discussion below Eq. (4)).
We have explored three types of $\rho_0(\epsilon)$: (i)a Gaussian (G),
which is not bounded, but can be interpreted as having an effective finite bandwidth, (ii) a semi-elliptic (S) form which has a compact support, and (iii) a wide, flat form (F). The expressions for the three forms are
\begin{align}
        \rho^{\rm G}_0(\epsilon) & =\frac{1}{\sqrt{\pi} t_*} \exp\!\left(-\frac{\epsilon^2}{t_*^2}\right)\, \\
    \rho_0^{\rm S}(\epsilon) & =\frac{1}{\pi t_*}\left(1-\frac{\epsilon^2}{4t_*^2}\right)^{1/2}\, \\
    \rho_0^{\rm F}(\epsilon) & = \frac{1}{2t_*}\theta(t_* - |\epsilon|) \,.
    \label{eq:flat}
\end{align}
We define an energy scale, $\Delta_0=\pi V_0^2\rho_0(0)$ which is known to  determine the scaling of dynamics and transport properties in equilibrium\cite{hewson_1993}, and hence can be expected to play an important role in the steady-state as well. In terms of this scale, the dot-lead coupling is given by, $\Gamma_{h\alpha}(\omega) = \Delta_0 \rho_0(\omega-\mu_\alpha - h\lambda)/\rho_0(0)$.
For all our calculations, we have used the semi-elliptic or the Gaussian forms with $t_*=1$, except for benchmarking where we have used the flat-DOS form with a large $t_*$, but a finite $\Delta_0$. In all the calculations described in this work, we have chosen $\Delta_0=0.1$, implying that the choice of $V_0$ is not the same for the three forms.
The infinitely wide flat-DOS hybridization is obtained by using the limit $t_*\rightarrow \infty$, and a concomitant scaling of $V_0\sim \sqrt{t_*}$, such that $\Delta_0=0.1$. Thus, in such a limit, the hybridization does not have a bias or spin-orbit coupling dependence while the semi-elliptic and the Gaussian forms do, and this difference leads to several observable consequences as we will discuss in Sec. III.

The dot Green's functions used in the current expression, given by Eq.~\ref{eqn:curr}, may be computed using the Dyson's equations given by~\cite{aligiamain}
\begin{align}
\label{eqn:dysons}
    [G^r_d(\omega)]^{-1} = [g^r_d(\omega)]^{-1}-\Sigma^r(\omega)\,, \\
    G^<_d(\omega) = |G^r_d(\omega)|^2\Bigg(\frac{g^<_d(\omega)}{|g^r_d(\omega)|^2}-\Sigma^<(\omega)\Bigg)\,,
\end{align}
where $g^r_d, g^<_d$ represent the non-interacting ($U=0$)
Green's functions of the dot, and $\Sigma^r, \Sigma^<$ are the retarded and lesser self-energies respectively. Obtaining the self-energies
represents the greatest challenge in computing the current. We have employed the interpolative perturbative approximation (IPA),  introduced by  Aligia\cite{aligiamain}, which is equivalent to the second-order Keldysh perturbation theory (KPT2) in the steady-state p-h symmetric limit, and to the iterative pertubration theory IPT\cite{kajueter1996new} in the equilibrium limit, to get the self-energies.  
The second-order expressions for the retarded and lesser self-energies are as follows\cite{aligiamain}:
\begin{multline}
\label{eq:secnd_ordr_sigma_r}
   \Sigma^{r}(\omega) = U^2\int \left(\prod_{i=1}^{3}d\epsilon_i D(\epsilon_i)\right)\, (\omega^+ +\epsilon_3-\epsilon_2-\epsilon_1)^{-1}\\
    \times \left[\tilde{f}(-\epsilon_1)\tilde{f}(-\epsilon_2)\tilde{f}(\epsilon_3) + \tilde{f}(\epsilon_1)\tilde{f}(\epsilon_2)\tilde{f}(-\epsilon_3)\right]
\end{multline}
and
\begin{multline}
\label{eq:secnd_ordr_sigma_lt}
    \Sigma^{<}(\omega) = -2i\pi U^2\int d\epsilon_1d\epsilon_2D(\epsilon_1)D(\epsilon_2)D(\epsilon_1+\epsilon_2-\omega)\\
    \times \left[\tilde{f}(\epsilon_1)\tilde{f}(\epsilon_2)\tilde{f}(\omega-\epsilon_1-\epsilon_2)\right]
\end{multline}
where $D(\omega)=-(1/\pi)\Im{\tilde{g}^r_d(\omega)}$ is the spectral
function calculated from the Hartree-corrected, retarded dot Green's function
given by
\begin{equation}
\label{eqn:hc_gfs}
    [\tilde{g}^r_{d\sigma}(\omega)]^{-1} = \omega^+  -\sum_{\alpha}\Delta_{\alpha}(\omega)\,,
\end{equation}
and $\tilde{f}(\omega)=\sum_\alpha \Gamma_\alpha (\omega) f_\alpha(\omega)/\sum_\alpha \Gamma_\alpha (\omega)$ is the weighted Fermi function, with $\Gamma_\alpha(\omega) = -{\rm Im} \Delta_\alpha(\omega)$ (Eq.~\ref{eq:hybksum}), and $f_\alpha(\omega) = f(\omega - \mu_\alpha)$ is the Fermi function for the $\alpha^{\rm{th}}$ lead.
In writing the above, we have assumed the p-h symmetric limit, where
$\epsilon_d=-U/2$ cancels the first order Hartree contribution.
To obtain the lesser Green's function [$G^<_d(\omega)$], we need the Hartree-corrected lesser Green's function of the dot, which is given by
\begin{equation}
\label{eqn:hc_gl}
    \tilde{g}^<_{d\sigma}(\omega) = 2i|\tilde{g}^r_{d\sigma}(\omega)|^2\sum_{\alpha}\Gamma_{\alpha}(\omega)f_\alpha(\omega)\,.
\end{equation}
Solutions of Eqs.~\ref{eqn:curr} to ~\ref{eqn:hc_gl} 
yield the physical picture of the interplay of bias, interactions and SOC on the spectra and current-voltage characteristics. The self-energies in Eq. \eqref{eq:secnd_ordr_sigma_r} and \eqref{eq:secnd_ordr_sigma_lt} are 
 evaluated as convolutions using the Fourier transform. The details of the numerical implementation of the convolution can be found in earlier works\cite{barman2011transport}. The differential conductance is calculated as $
    G = dJ/dV_{sd}$, and numerically implemented through a derivative of the splined current\footnote{Codes for reproducing the IPA data in Fig. 2 and Fig. 6 are provided at the GitHub link (\url{https://github.com/nsvraja/ssIPA_ZF_phsym}).  A README is also included for ease of use.}.
We present our results for spectra and transport quantities in the next section.

\section{Results}

In equilibrium studies of the p-h symmetric Anderson model, the IPT\cite{kajueter1996new}, based on the second-order perturbation theory, is a good approximation in the weak coupling limit, and by coincidence also reproduces the atomic limit. Hence, the IPT has been extensively used as an interpolating approximation for lattice models such as the Hubbard model and the periodic Anderson model within dynamical mean-field theory to investigate Mott transition and heavy fermion physics\cite{georges1996dynamical}. Various Keldysh perturbation theory based approximations including IPA have been used quite widely in the out-of-equilibrium case as well \cite{joura2015long,fujii2003perturbative,aligiamain}. In this work, we carry out a simple benchmarking exercise to ascertain the regime of validity of the IPA and subsequently use it to investigate the effect of spin-orbit coupling.

\subsection{Benchmarking}

\begin{figure}[thb]
    \centering
    \includegraphics[width=\linewidth]{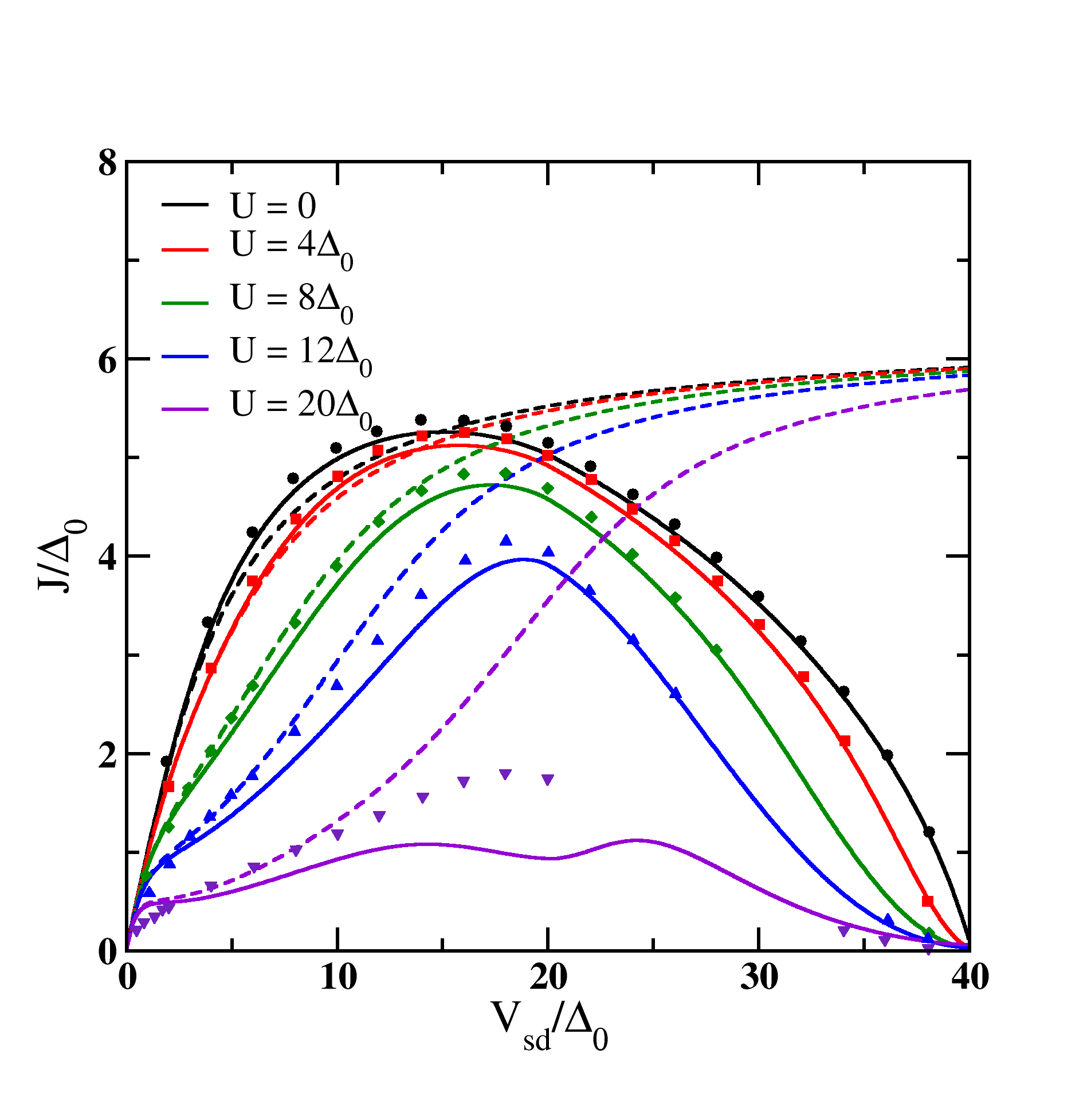}
    \caption{The current-voltage curves computed within IPA ($\lambda=0$) for various values of $U$ (mentioned as legends) compared to TEBD results of Nuss~\etal \cite{nuss2013steady}. The solid lines represent the current computed for a semi-elliptic hybridization, while the dashed lines correspond to the flat hybridization. The symbols are data extracted from Nuss \etal\cite{nuss2013steady} and the current is scaled by a factor of 3.33.}
    \label{fig:tebd_benchmark}
    \end{figure}

    As a first benchmark, we compare the current-voltage (IV) curves for the flat-DOS and the semi-elliptic DOS cases with the steady state results from Nuss \etal\cite{nuss2013steady} who have used the time-evolving block decimation (TEBD) method on a model with 150 sites comprising two leads and a quantum dot subjected to a dc bias. As shown in Fig. \ref{fig:tebd_benchmark} we find that the $J-V_{sd}$ curves match the TEBD results very well\footnote{For a fixed energy scale $\Delta_0$, the hybridisation strength $V_0$, needs to have a multiplicative factor of $\sqrt{2}$ for a quantitative comparison with the TEBD results}. The presence of a peak in the current at a specific bias roughly around half the bandwidth is seen in the case of the finite bandwidth semi-elliptic DOS case. The peak position moves to 
    higher bias values for increasing interaction strength and concomitantly the peak current magnitude decreases. In the case of the infinitely wide, flat hybridization function, we see a saturation of the current
    at high bias values and the saturation current decreases with increasing $U$. Both of these results show that the IPA captures the current characteristics very well over a wide range of interaction strengths.
    
   \begin{figure}[thb]
    \centering
    \includegraphics[width=\linewidth]{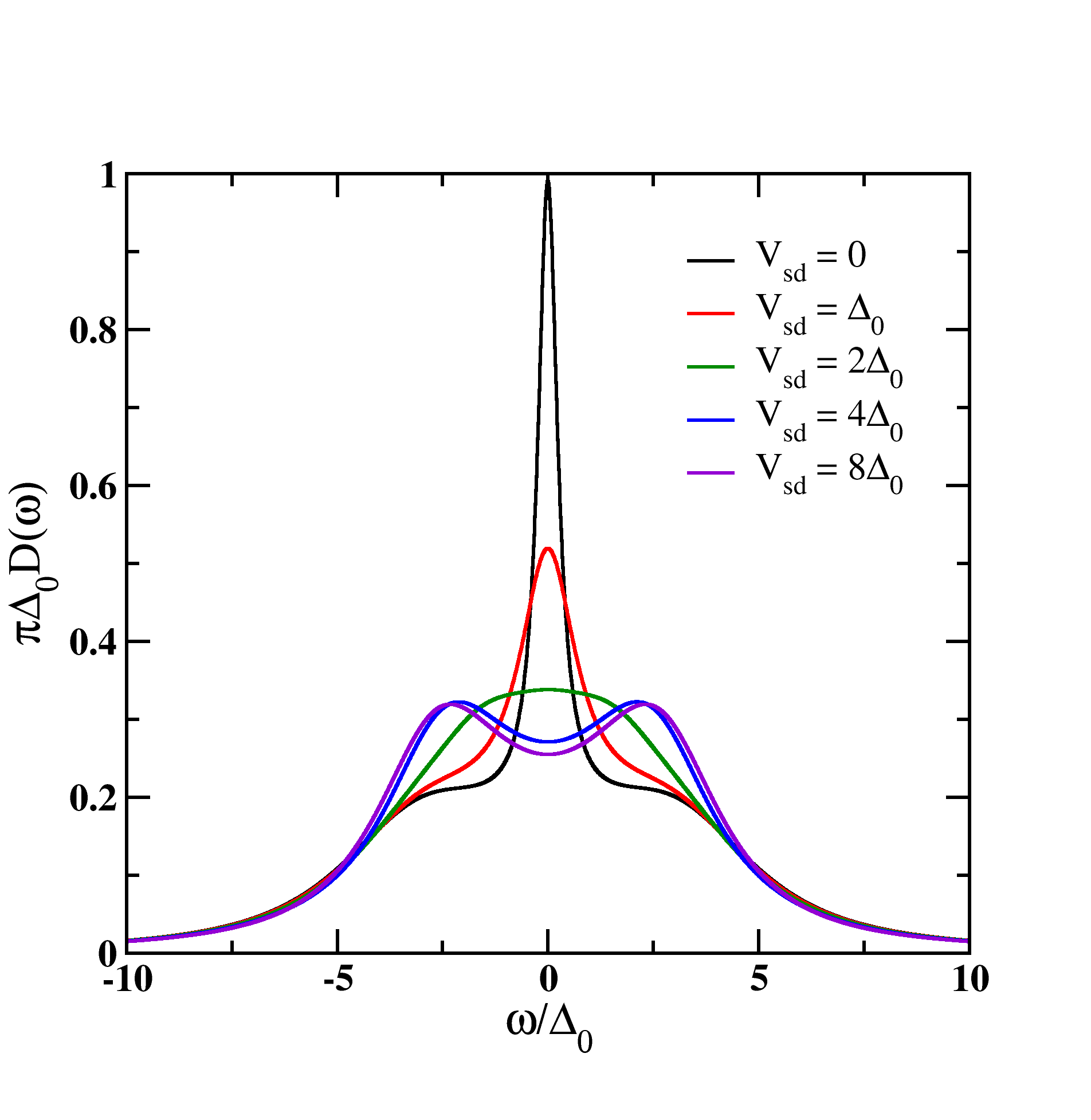} 
    \caption{The DOS of the dot computed within IPA for $U=6\Delta_0$ and the flat hybridization
    in the absence of SOC. A splitting of the zero frequency peak with increasing bias (indicated as legend) is seen, which matches with the results of a fourth-order calculation of Fujii~\etal\cite{fujii2003perturbative}.
    }
    \label{fig:fujii_bench}
    \end{figure}    
    As a second benchmark, we compare our results with fourth-order perturbation theory by Fujii \etal\ \cite{fujii2003perturbative} (for a flat/uniform hybridization, with $U=6\Delta_0$ and SOC strength $\lambda=0$) in Fig. \ref{fig:fujii_bench} and see that the interacting dot DOS matches very well with their results.  We see that there is a reduction in the zero frequency weight with increasing bias and also splitting and broadening of the central peak. Next, we examine the effect of equilibrium scales on transport characteristics in the linear response regime.

    \subsection{Linear response regime: scaling and universality}
     
The excellent agreement with TEBD and fourth-order perturbation theory yield strong confidence in the validity of IPA in a wide bias and interaction range. We now explore the extent of the linear response regime and examine if equilibrium quantities describe the response of the system at finite bias, but close to equilibrium. 

 As we can see from Fig. \ref{fig:tebd_benchmark}, the linear dependence of the current at low values of bias agrees well with the TEBD calculations and the extent of the linear response regime depends on the value of the interaction and SOC strengths. In equilibrium, the quasiparticle weight, which can be calculated as $Z_0= [1- {\rm Re}(d\Sigma(\omega)/d\omega|_{\omega=0})]^{-1}$, where $\Sigma(\omega)$ is the self energy, happens to be proportional to the Kondo scale, and hence determines the extent of the universal regime\cite{hewson_1993}. In order to understand the dependence of the self-energy on SOC and bias, we have carried out a simple analysis (details in the Appendix) of the second-order expression (Eq.~\ref{eq:secnd_ordr_sigma_r}) assuming that the Hartree-corrected density of states is not varying rapidly in the neighbourhood of $\omega=0$ for low values of bias, and SOC, i.e $D(\omega)\simeq D_0=-\Im{\tilde{g}^r_{d\sigma}(\omega=0)}$ for $|\omega|,|V_{sd}|,\lambda << t_*$. $D_0$ may be found by considering the expressions for the retarded Green's functions
 in Sec. II, as shown below.
From Eq.~\ref{eqn:hc_gfs}, we see that 
\begin{equation}
    \left[\tilde{g}^r_{d\sigma}(\omega=0)\right]^{-1} = V_0^2\sum_{\alpha h}
    \int\,d\epsilon\,\frac{\rho_0(\epsilon)}{\mu_\alpha+h\lambda+\epsilon - i\eta}\,
\end{equation}
where $\eta\rightarrow 0^+$. Using the above expression, we note that 
at $\omega=0$, the Hartree-corrected non-interacting Green's function
$\tilde{g}^r_{d\sigma}$ is purely imaginary, since
\begin{equation}
    {\rm Re} \left(\tilde{g}^r_{d\sigma}(\omega=0)\right)^{-1} = V_0^2\sum_{\alpha h}
{\cal{P}}\int\,d\epsilon\,\frac{\rho_0(\epsilon)}{\mu_\alpha+h\lambda+\epsilon}
=0\,
\end{equation}
for a p-h symmetric $\rho_0(\epsilon)$. In the above equations,  $\mu_L=V_{sd}/2, \mu_R=-V_{sd}/2, h=\pm 1$, and ${\cal{P}}$ denotes the principal value. Thus, 
\begin{equation}
   \tilde{g}^r_{d\sigma}(\omega=0) = -\frac{i}{\pi V_0^2\sum_{h\alpha}\rho_0(\mu_\alpha + h\lambda)}\,
\end{equation}
using which we can write, for zero bias, $D_0=(4\pi^2V_0^2\rho_0(\lambda))^{-1}$.
We find that the imaginary part of the retarded self-energy in terms of $D_0$ is given by (for $\omega\rightarrow 0$ and $T=0$, details are provided in the Appendix)
\begin{equation}
    -\frac{1}{\pi}\Im\Sigma^r(\omega) = \frac{U^2 D^3_0}{2}\left[\omega^2+\frac{3V_{sd}^2}{4}\right]
    \label{eqn:imse}
\end{equation}
Using the Kramers-Kr\"{o}nig transformation, the real part of the retarded self-energy is given by
\begin{equation}
    \Re\Sigma^r(\omega) = -\frac{U^2 D^3_0}{\Lambda}\left[\Lambda^2-\frac{3V_{sd}^2}{4}\right]\omega\,,
\end{equation}
where $\Lambda$ is a high-energy cutoff that represents the extent of the quadratic dependence of the imaginary part of the self-energy. 

Thus, using the definition of the quasiparticle weight as $\Re\Sigma^r(\omega) = \omega(1-\frac{1}{Z}) $ we get the analytical expression (please refer to the Appendix for details of the calculation),
\begin{equation}
    Z(U,\lambda,V_{sd}) = \left(1+\frac{U^2 D^3_0}{\Lambda}\left[\Lambda^2-\frac{3V_{sd}^2}{4}\right]\right)^{-1}\,.
    \label{eqn:zmain}
\end{equation}

\begin{figure}[thb]
        \centering
        \includegraphics[width=0.99\linewidth]{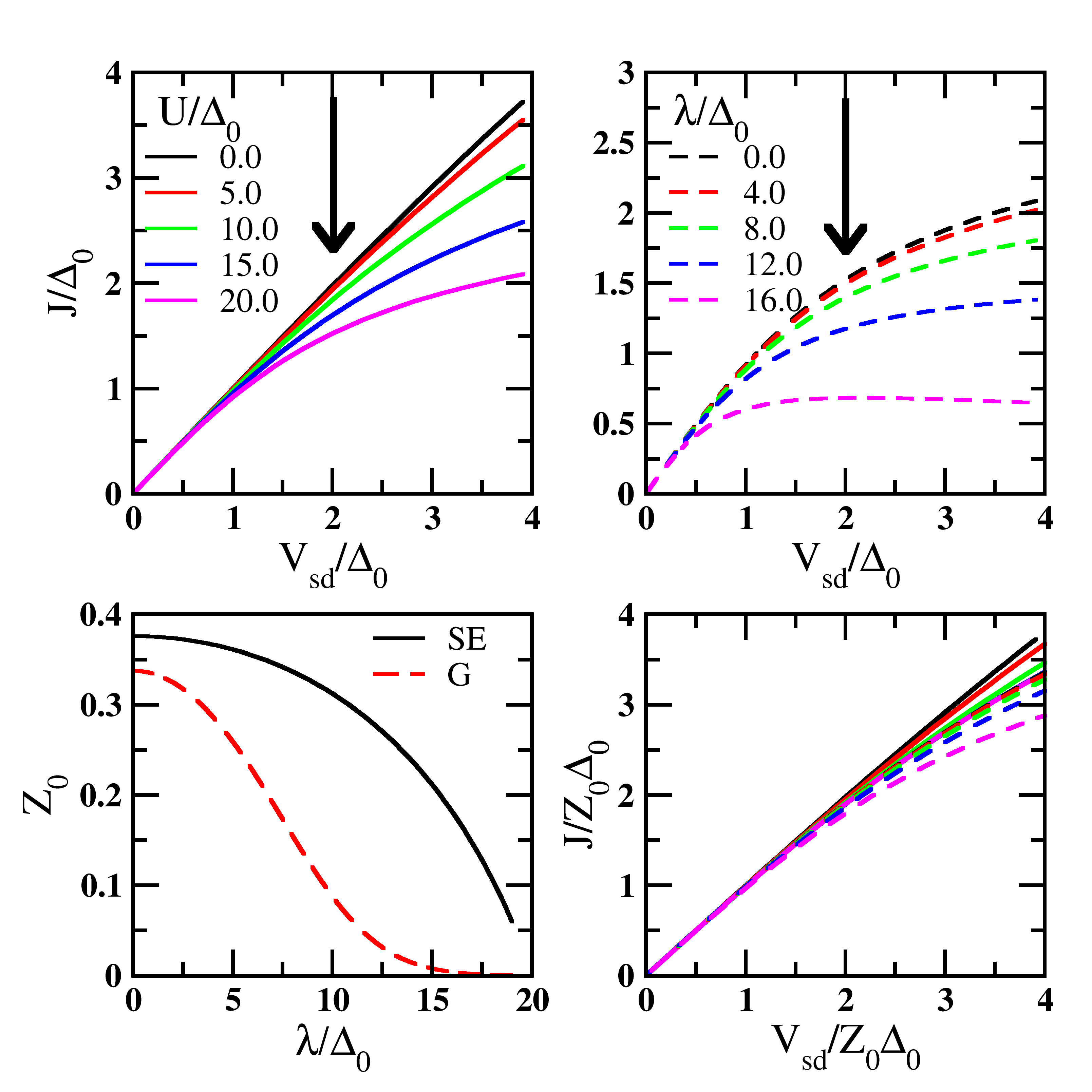}
        \caption{Current ($J/\Delta_0$) as a function of bias ($V_{sd}/\Delta_0$) is shown in the top two panels. The left panel shows the result for various values of interaction strength $U/\Delta_0$, and SOC strength $\lambda=0$, while the right panel shows the result for $U=20\Delta_0$ and various values of $\lambda/\Delta_0$. The bottom left panel shows the dependence of equilibrium ($V_{sd}=0$) quasiparticle weight $Z_0$ on the spin-orbit coupling for the semi-elliptic (SE) and Gaussian (G) forms of hybridization. The bottom right panel shows the collapse of all the curves in the top two panels up to a bias given by $V_{sd}\simeq Z_0\Delta_0$, when the current and bias are scaled by $Z_0$, thus determining the linear response regime. }
        \label{fig:scaled_int}
    \end{figure}

Using this, we note that, since the IPA is based on second-order perturbation theory, the equilibrium quasiparticle weight, $Z_0=Z(U,\lambda,V_{sd}=0)$ decays algebraically with increasing 
interaction strength (at zero bias) as $U^{-2}$. The dependence on SOC enters through $D_0$. For a flat density of states, since $D_0$ does not depend on $\lambda$, the quasiparticle weight, $Z_0$ will be independent of SOC. The Kondo scale is proportional to the product of bandwidth and $Z_0$, and since the chiral bands move outward with increasing $\lambda$, the effective bandwidth will increase and hence will give rise to a linearly increasing Kondo scale. However, for a frequency-dependent density of states such as the semi-elliptic or Gaussian density of states, $D_0$ increases as $\rho_0^{-3}(\lambda)$. For example, for the Gaussian and semi-elliptic forms of the hybridization, the $D_0$ is proportional to $\exp(3\lambda^2/t_*^2)$ and $(1-\lambda^2/4t_*^2)^{-3/2}$ respectively. Hence the quasiparticle weight will decrease sharply with increasing $\lambda$ as $Z_0\propto D_0^{-3}$ [from Eq.~\ref{eqn:zmain}]. So, even though the effective bandwidth increases linearly with increasing SOC in parallel to the flat hybridization case, the strong decrease due to the factor of $D_0^{-3}$ dominates. The equilibrium quasiparticle weight, $Z_0$ shown in the bottom left panel of Fig.~\ref{fig:scaled_int}, for the semi-elliptic and the Gaussian hybridization functions (denoted as SE and G respectively), is seen to decrease sharply with increasing $\lambda/\Delta_0$ in agreement with the arguments above. 

Consistent with the decrease of $Z_0$ with $\lambda$, the linear response regime in the current-voltage ($J-V_{sd}$) relationship shrinks considerably with increasing interaction strength (for $\lambda=0$) as seen in the top left panel of Fig.~\ref{fig:scaled_int} and with increasing $\lambda$ for $U=20\Delta_0$ as seen in the top right panel. However, when the current and bias are scaled by $Z_0$, all the curves of the top two panels collapse up to a bias of $V_{sd}\simeq Z_0\Delta_0$ (shown in the bottom right panel), which confirms that the equilibrium quasiparticle weight, or equivalently, $T_K$, determines the extent of the linear response regime~\cite{grobis2008universal}.

Equilibrium investigations of the SIAM have shown that the Kondo scale decreases exponentially with increasing $U$ for $\lambda=0$~\cite{tsvelick1983exact}.
In the presence of $\lambda$ on the leads and $U$ on the dot, the Kondo scale has been shown to increase or decrease depending on whether the leads have a wide, flat form or a finite bandwidth\cite{sandler,bonvca2011kondo} respectively. In our work, we have shown [see Eq.~\ref{eqn:zmain}] that the quasiparticle weight, and hence the Kondo scale depends on $U$ algebraically (decreasing as $U^{-2}$) and on $\lambda$ through 
the density of states, namely as $(\rho_0(\lambda))^3$.
The algebraic decrease of the scale with $U$ is known to be an artifact of the perturbative approximation employed to obtain the self-energy, while exact methods such as Bethe ansatz\cite{tsvelick1983exact} find an exponential decrease. Since $\rho_0(\epsilon)$ is usually chosen to have a maximum at the chemical potential, and decreases monotonically with increasing $|\epsilon|$, the Kondo scale will also decrease monotonically with increasing $\lambda$ as discussed above. The precise form of this decrease will depend on the form of the density of states.
We see from Fig.~\ref{fig:scaled_int}, (bottom-left panel) that the quasiparticle weight scale decreases sharply with increasing $\lambda$. Although the qualitative trend is the same for the semi-elliptic and the Gaussian hybridization, the scale decreases much more rapidly for the latter. For the same value of the parameters, the scale corresponding to the Gaussian hybridization is much smaller than that of the semi-elliptic case. Hence for the strong coupling regime, we choose to work with the Gaussian hybridization.

Additionally, in the strong coupling\footnote{In practice, we consider the system to be in the strong coupling regime when the $Z_0 \lesssim 0.1$, since the low-temperature scale, i.e $T_K\sim Z_0\Delta_0$ will be at least one order of magnitude smaller than the non-universal scales, $\Delta_0$, bandwidth, $U$, etc.} ($T_K\rightarrow 0$) regime, a clear separation of scales ($T_K$ {\it vs} bandwidth, $U$) occurs, which is the best distinguishing feature of this regime, and will be seen to have important consequences on the evolution of spectra and conductance with increasing temperature, and will be discussed later.

Subsequent to the benchmarking and a study of the scales and the linear response regime, we now present the main results of our work, which focus on the interplay of spin-orbit coupling and electron-electron interactions on the single-particle spectra and the differential conductance. Since the SOC does not alter the hybridization functions in the case of the infinitely wide, flat/uniform DOS, the results do not show any dependence on $\lambda$. Therefore, we will consider the  semi-elliptic and Gaussian hybridization forms where the effective bandwidth is modified by $\lambda$ as seen from Eqs.~\ref{eq:dh}-~\ref{eq:flat}. 

\subsection{Evolution of density of states with interactions, SOC, bias, and temperature}
Since the separation of scales is an important consideration in our analysis, we will investigate the transfer of spectral weight in the weak/intermediate coupling and strong coupling regimes separately.

\subsubsection{Weak/intermediate coupling regime}
\begin{figure}
       \centering
       \includegraphics[width=0.99\linewidth]{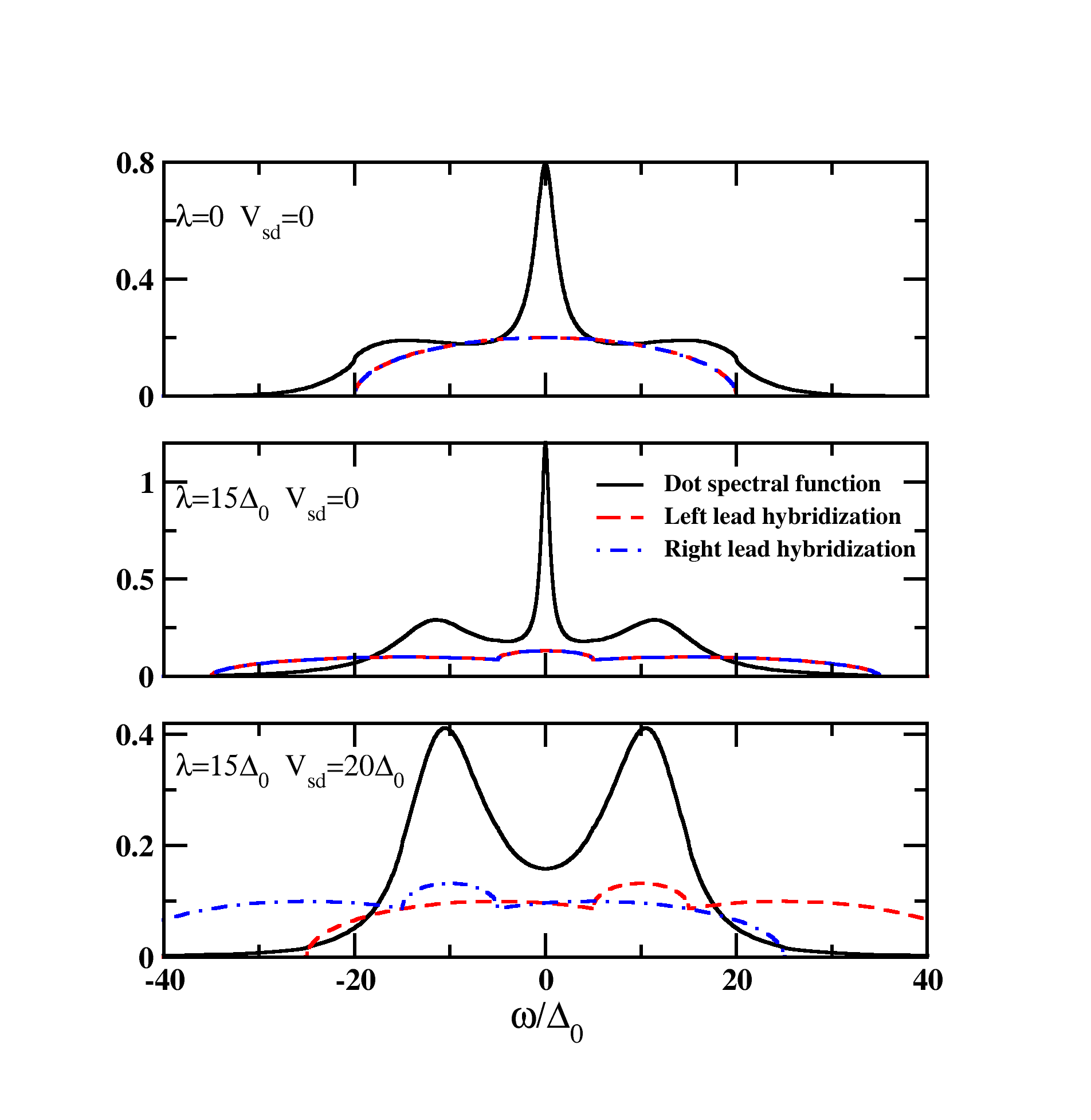}
       \caption{The dot spectral function (solid black line), and the left and right lead hybridization functions (red dashed and blue dot-dashed lines respectively) as a function of $\omega/\Delta_0$ showing the variation of spectral features with increasing SOC and bias for a fixed interaction strength, $U=20\Delta_0$. The top panel is for $\lambda=V_{sd}=0$, while $\lambda=15\Delta_0$ is turned on in the middle panel keeping bias zero, and the bottom panel has $\lambda=15\Delta_0$ and $V_{sd}=20\Delta_0$. The semi-elliptic hybridization function has been used here.}
       \label{fig:fin_bias_pk}
   \end{figure}

   In order to be in the weak/intermediate coupling regime, we will choose $U=20\Delta_0, \lambda=15\Delta_0$ and the semi-elliptic hybridization, for which, as the bottom left panel of Fig.~\ref{fig:scaled_int} shows, the $Z_0\simeq 0.2$.
    At equilibrium (zero bias), and in the absence of SOC, but with finite $U$ ($\gtrsim$ bandwidth), the dot spectral function acquires Hubbard bands at $\omega\gtrsim \pm U/2$ as the top panel of Fig. \ref{fig:fin_bias_pk} shows. For a flat/uniform hybridization, the Hubbard bands are known to lie close to $\pm U/2$\cite{syaina2018theoretical}, but for a dispersive DOS such as semi-elliptic or Gaussian, these incoherent peaks lie somewhat beyond $\pm U/2$. If we now turn on SOC, keeping $V_{sd}=0$, then as the middle panel shows, the left, and right lead hybridization functions broaden significantly. This is because the chiral bands that are split by the SOC, when superimposed give rise to a hybridization, that has a width equal to $D+2\lambda$, and for the middle panel, since $D=40\Delta_0$, and $\lambda=15\Delta_0$, the band-edges are at $\pm(D/2+\lambda)=\pm 35\Delta_0$. Since, for this larger bandwidth, the hybridization appears less dispersive and appears similar to a uniform DOS, the Hubbard bands become more prominent, and their location is almost at $\pm U/2$. 
    
    Next, as bias is turned on and raised to $20\Delta_0$, the band centers (chemical potentials) of the two leads move apart ($\mu_L=V_{sd}/2$ and $\mu_R=-V_{sd}/2$), and concomitantly, the dot spectrum (solid black line in the bottom panel of Fig.~\ref{fig:fin_bias_pk}) goes over to a two-peak structure in non-equilibrium from a three-peak structure at equilibrium. In the same panel, the red dashed and blue dot-dashed lines, which are the left and right lead hybridizations, have small bumps at precisely these energies for $V_{sd}=20\Delta_0$.
    Since the Hubbard bands are also at $\pm U/2$, which are
    $\pm 10\Delta_0$ in Fig.~\ref{fig:fin_bias_pk}, a maximum
    in conductance may be expected to occur when the bias becomes equal to the 
    peak position difference of the Hubbard bands~\cite{fujii2003perturbative}. Indeed, this will be confirmed in Sec. III D-1. 

\subsubsection{Strong coupling regime}
In parallel to the previous section, we will analyze the changes in the spectra and hybridization as we sequentially turn on SOC, temperature, and bias for $U=20\Delta_0$. The strong coupling regime will be accessed through a choice of the value of the SOC, and the hybridization function, as
$\lambda=18\Delta_0$, and the Gaussian, for which as the left bottom panel of Fig.~\ref{fig:scaled_int} shows, the scale is $Z_0\sim 10^{-3}$.
 \begin{figure}
       \centering
    \includegraphics[scale=0.625]{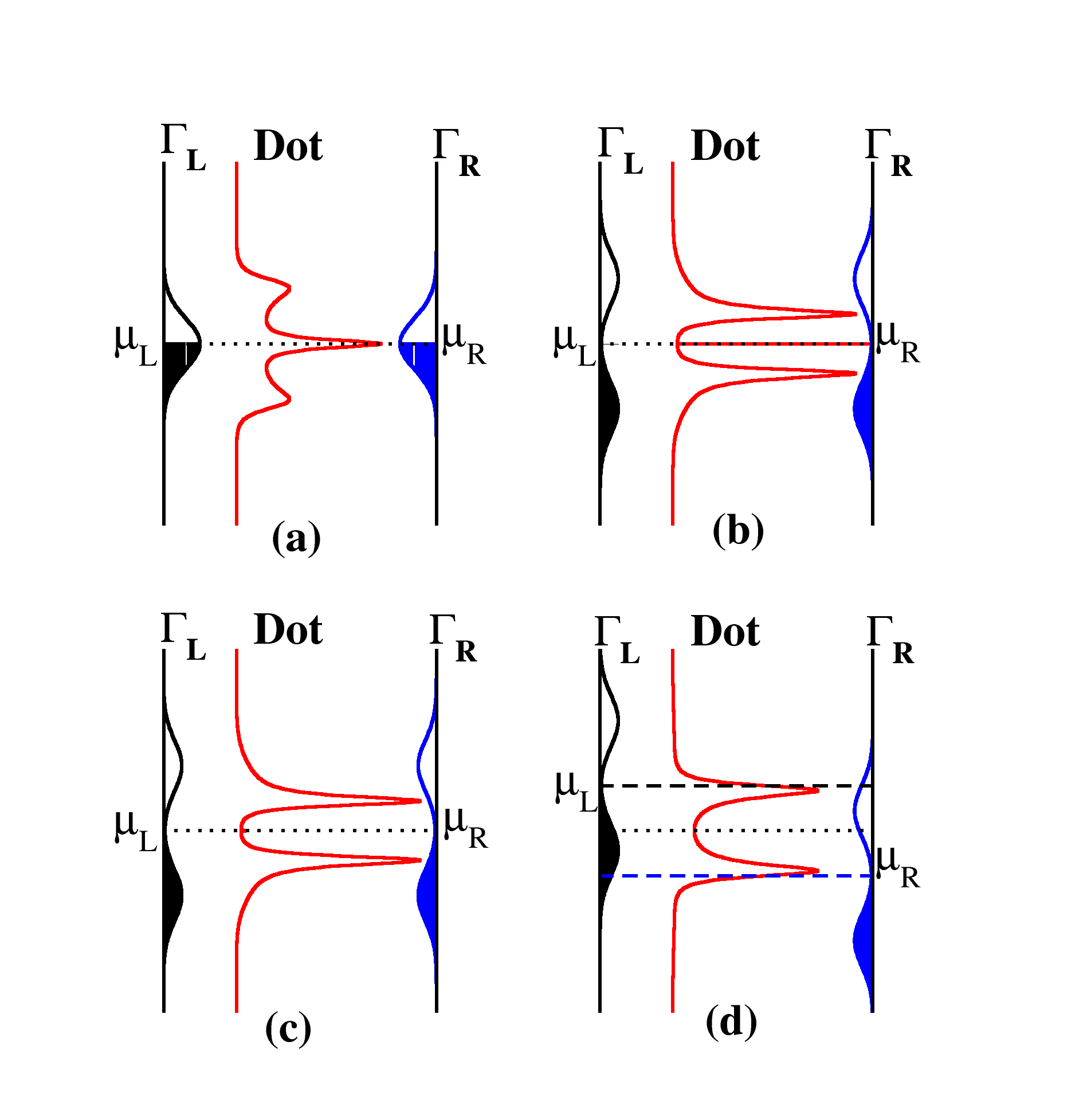}
       \caption{The hybridization and the dot density of states, for the Gaussian hybridization function in the strong coupling regime. $\Gamma_\alpha$ is the lead hybridization with $\alpha=L/R$ for left (black)/right (blue), and the central (red) curve shows the dot density of states. Similarly, $\mu_\alpha$ represents the chemical potential of the $\alpha^{\rm th}$ lead. The interaction strength is fixed at $U=20\Delta_0$, while SOC, temperature and bias are turned on sequentially from (a)-(d) as follows: (a) $\lambda=T=V_{sd}=0$; (b) $\lambda=18\Delta_0$, $T=V_{sd}=0$; (c) $\lambda=18\Delta_0, T=0.05\Delta_0$, $V_{sd}=0$; (d)$\lambda=18\Delta_0, T=0.05\Delta_0$, $V_{sd}=25\Delta_0$. The shaded regions represent the $T=0$ occupied states. The dotted and dashed lines are guides to the eye.}
       \label{fig:schem}
   \end{figure}

Fig.~\ref{fig:schem}(please see footnote\footnote{The figure has been made to appear as a schematic, but it has been made using data obtained from IPA, and the curves in the four panels are not mere sketches. The purpose of this figure is to allow a direct comparison of the calculated results with the schematic shown in Fig. 1 of Ref~\cite{smith2022electrically}}) shows the left lead and the right lead hybridization functions as $\Gamma_L$ and $\Gamma_R$ in black and blue colors respectively. The dot spectral function is shown in the center as a red solid line. The vertical axis is energy/frequency.  Figure 6(a) represents a situation where $U=20\Delta_0$, but $\lambda=T=V_{sd}=0$. The dot spectral function has a three-peak structure, with the central peak being the Kondo peak, and the other two being the Hubbard bands at $\pm U/2$. As we turn on $\lambda=18\Delta_0$, keeping $T=V_{sd}=0$,
Fig. 6(b) shows that the central peak in the dot spectral function becomes extremely narrow (width$\simeq Z_0\Delta_0
\sim 10^{-4}$), while $\Gamma_L$ and $\Gamma_R$ show the development of chiral bands due to the SOC. The clear separation of the Kondo peak and the Hubbard bands is evident, and is a characteristic of the strong coupling regime. When we turn on a small, but finite temperature
 of $T=0.05\Delta_0$, Fig. 6(c) shows that the central peak melts leaving the Hubbard bands as the only distinct features in the spectrum. Finally, when a bias of $V_{sd}=25\Delta_0$ is turned on, there are minor changes in the positions of the spectral peaks, while the $\Gamma_L$ and $\Gamma_R$ move up and down respectively. The occupied region of the left lead is seen to overlap with the region between the Hubbard bands, and the unoccupied region of the right lead. This overlap can very likely lead to finite bias peaks in the conductance.  Next we analyze the dependence of the differential conductance on bias, SOC, interactions and temperature.

    \begin{figure*}
            \centering
            \includegraphics[width=0.49\linewidth]{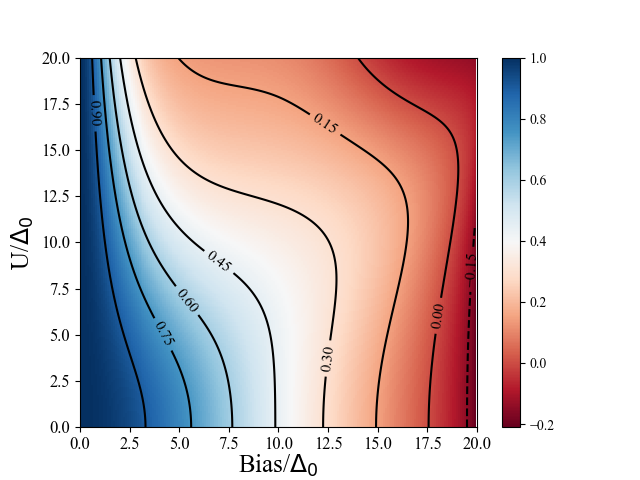}
            \includegraphics[width=0.49\linewidth]{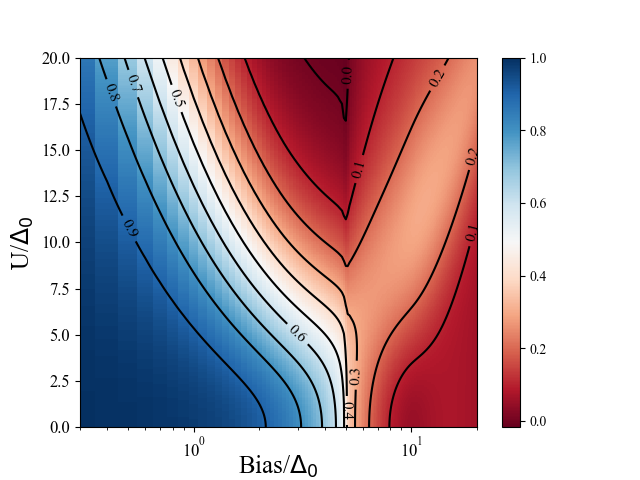}
            \includegraphics[width=0.49\linewidth]{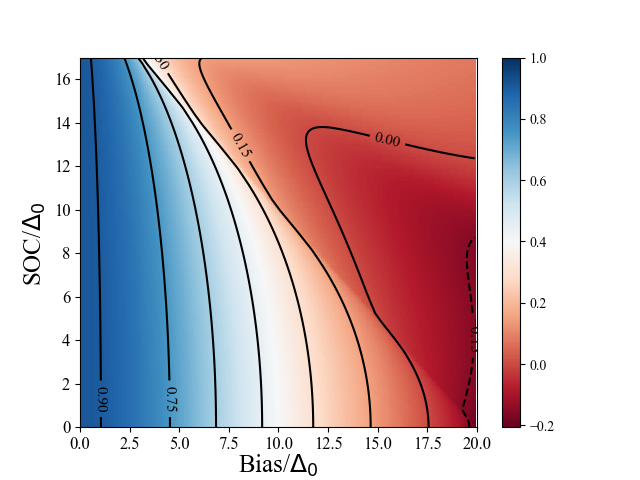}
            \includegraphics[width=0.49\linewidth]{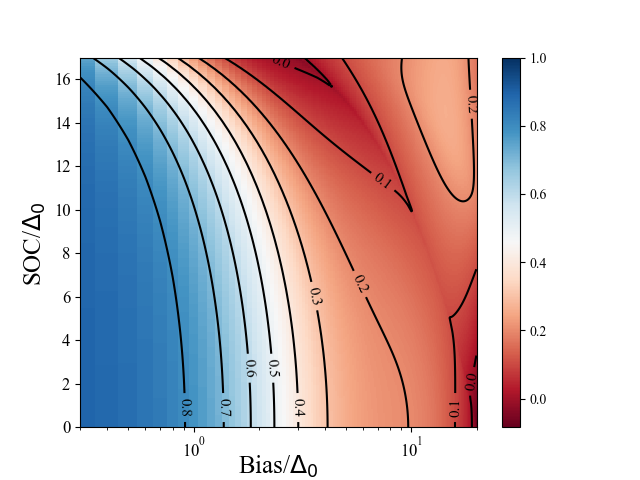}
            \caption{Variation of the differential conductance with interaction ($U$), spin-orbit coupling (SOC) and bias. The colors correspond to the values of the conductance $G=dJ/dV_{sd}$ as mentioned in the color bar with certain values marked by the black contours. The dashed lines represent the negative differential resistance regime The top left and right panels correspond to varying $U$ and fixed SOC equal to zero and $15\Delta_0$ respectively. The bottom left and right panels correspond to varying SOC and fixed $U$ equal to zero and
            $15\Delta_0$ respectively. Note that the top and bottom right panels show the $x$-axis on the logarithmic scale for clarity, since the linear response regime is highly compressed for large $U$ and/or SOC.}
            \label{fig:conduct_lamb}
    \end{figure*}

\subsection{Conductance}
    
    We have divided the conductance results into two sections. In the first section, we present the differential conductance in the weak/intermediate coupling regime focusing on the evolution of various features with $U, \lambda$, SOC and $T$.  
    The second section focuses on the strong correlation regime and qualitative comparison to experimental results.

    \subsubsection{Weak/Intermediate coupling regime}
    Figure \ref{fig:conduct_lamb} shows color contour plots of the conductance for various values of $U\in [0,20]\Delta_0$ and $\lambda\in[0,15]\Delta_0$ at $T=0$, computed with the semi-elliptic hybridization function.
    Within this range of parameters, the quasiparticle weight
    is in the range of $Z_0\in [0.2,1]$, so the choice of the range of parameters corresponds to the weak/intermediate coupling regime.
    The conductance is computed by fitting a cubic spline to the current {\it vs} $V_{sd}$ data and taking the first derivative. The color bar on the right of each panel shows the association of a given color with the value of the conductance. The top left and right panels represent varying interaction strength on the y-axis, but fixed SOC of zero and $15\Delta_0$ respectively, while the bottom left and right panels represent varying SOC on the $y$-axis, but fixed $U$ of zero and $15\Delta_0$ respectively. The ZBC peak is seen to be a universal feature in all panels, and it is interesting to note that the width of this peak decreases with increasing $U$ with or without $\lambda$, while the effect of $\lambda$ on the peak width at $U=0$ is very weak (bottom left panel), but in the presence of $U$, $\lambda$ leads to a sharp narrowing of the zero bias peak (bottom right panel), which is consistent with the bottom left panel of Fig.~\ref{fig:scaled_int}.
    
    With increasing bias, a negative differential conductance (NDC) regime is seen in all panels. In order to understand the origin of the NDC, we first focus on the $U=0$ results, since the calculation of conductance for $U=0$ does not depend on the second-order approximation for the self-energy, and is hence exact. The non-interacting regime also exhibits NDC at high bias values as seen in the top panels and the bottom left panel. 
    We confirm that the finiteness of the bandwidth of the hybridization is responsible for the NDC\cite{evlashin2021negative} since the wide, flat hybridization does not yield NDC. Additionally, the top right panel shows an intermediate bias regime where NDC is obtained at large $U$ values. In the strong coupling regime, reached by increasing $U$ and/or $\lambda$, the width of the central Kondo peak in the dot DOS, being proportional to $T_K$, shrinks exponentially. However, adiabatic continuity to the non-interacting limit ensures that the Kondo peak has precisely the same form as the non-interacting dot DOS
    if the frequency is scaled as $\omega/Z_0\Delta_0$\cite{logan1998local}. This fact implies that the system has an effective bandwidth of $Z_0\Delta_0$ in the strong coupling regime, and hence as the bias value crosses the linear response regime, an NDC regime may be obtained in parallel to the non-interacting regime. Since IPA is perturbative, and not accurate in strong coupling the NDC seen in strong coupling at intermediate bias values could be due to both the finite support of the hybridization and the approximation used for computing the self-energy.

    The top left panel of Fig. \ref{fig:conduct_lamb} shows that in the absence of SOC, the conductance decreases monotonically with increasing bias, and the linear response regime (over which the conductance is close to 1) shrinks with increasing interaction strength. The top right panel shows that in the presence of strong SOC ($15\Delta_0$), the conductance becomes highly non-monotonic, and a light band (for $U\gtrsim 10\Delta_0$) appears signifying a peak at a non-zero bias, that shifts to higher bias with increasing interaction strength. In fact, we observe that the bias value at which the peak occurs is almost the same as the interaction strength. Another interesting observation can be made from the bottom right panel of Fig. ~\ref{fig:conduct_lamb}, which shows the existence of a finite bias conductance peak for $\lambda\gtrsim 10\Delta_0$, whose position (on the bias-axis) is almost independent of the value of SOC. Thus, we see that the position of the finite bias peak correlates very well with the interaction strength, but is independent of $\lambda$, which suggests that the Hubbard bands and their increasing prominence with increasing $\lambda$ are giving rise to this peak. A physical basis for the emergence of this finite bias conductance peak in terms of the spectral functions and the hybridization functions may be constructed using the results of Sec. C-1.

    As Fig.~\ref{fig:fin_bias_pk} shows, an argument for the conductance peak at a bias value equal to interaction, is that the chemical potential of the left lead matches with one Hubbard band, while that of the right lead matches with the other, and hence the conductance is peaked due to a resonant situation when the SOC is large ($\gtrsim 10 \Delta_0$), as the bottom right panel of Fig.~\ref{fig:conduct_lamb} shows. Indeed, the left bottom panel, for which $U=0$, shows that the conductance decreases monotonically with increasing bias, while the bottom right panel shows  a peak at bias around $15\Delta_0$ which is present only for large SOC ($\gtrsim 10\Delta_0$), but independent of the value of SOC. Thus, we find that, at $T=0$, in addition to the zero bias peak, a finite bias peak arises in the presence of strong spin-orbit coupling, which is positioned at a bias roughly equal to the value of the interaction strength.

    \begin{figure}
       \includegraphics[width=1.05\linewidth]{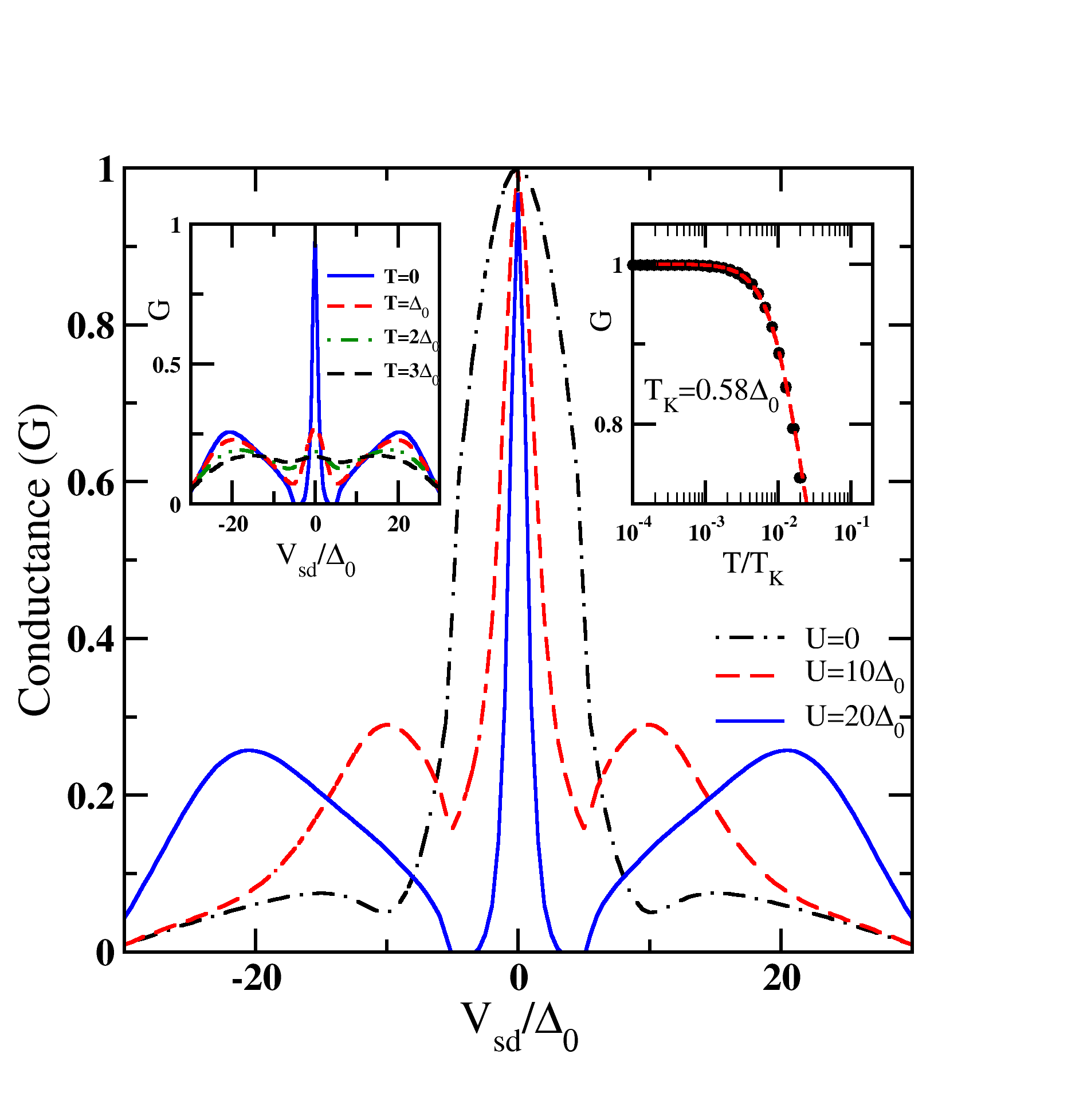}
       \caption{The main panel shows the $T=0$ differential conductance in units of $G_0$ as a function of scaled bias ($V_{sd}/\Delta_0$) for three values of interaction strength, namely $U=0$ (dot-dashed line), $10\Delta_0$ (dashed line), $20\Delta_0$ (solid line). The SOC is fixed at $\lambda=15\Delta_0$ and a semi-elliptic hybridization function has been used in the calculations. The left inset shows $G$ {\it vs} $V_{sd}/\Delta_0$ for $U=20\Delta_0$ at four different temperatures, mentioned as legends. The central peak is seen to melt with increasing temperature. The right inset shows the ZBC as a function of $T/T_K$; The black symbols represent IPA data, and the red dashed line represents a fit, $G(T)=G_0(1+(2^{1/s}-1)(T/T_K)^2)^{-s}$ with $s=0.22$ and $T_K=0.58\Delta_0$ as appropriate for universal Kondo behavior}
       \label{fig:finT}
   \end{figure}

   We now focus on finite temperature effects on conductance and also examine whether the IPA results conform to universal Kondo scaling of the ZBC. We extend the bias to negative values, and show the conductance at zero temperature and a fixed SOC of $\lambda=15\Delta_0$ as a function of bias, for various $U$ values, and a semi-elliptic hybridization function, in the main panel of Fig. ~\ref{fig:finT}.
   The conductance at $T=0$ is seen to have a three-peak form consistent with the p-h symmetric regime, and the results of Fig.~\ref{fig:conduct_lamb}. For the finite $U$ values, the locations of the side peaks correspond closely to the value of $U/\Delta_0$, while the zero bias peak is characteristic of the linear response regime. We can also see that the full width at half maximum (FWHM) of the central peak decreases sharply with increasing interaction, while the conductance at zero bias is pinned at unity. 
   The width of the central peak is roughly proportional to the zero bias quasiparticle weight\cite{kretinin2011spin}, which decreases exponentially (algebraically within IPA) with increasing $U/\Delta_0$. The left inset of Fig.~\ref{fig:finT} shows the temperature dependence of the conductance for the largest $U/\Delta_0$ considered in the main panel. Interestingly, the central peak melts, and thus, at modest temperatures, we see a two-peak structure, which resembles the results of a 
   recent experimental study~\cite{smith2022electrically}. However, the right inset shows that the conductance at zero bias decreases monotonically with increasing temperature in accordance with the universal Kondo behavior~\cite{goldhaber1998kondo}, while the `double zero bias peak' feature seen in experiments~\cite{smith2022electrically} exhibits the opposite behavior, namely an increase of $G$ with temperature. A rise in the ZBC is indeed observed at higher temperatures, but the rise is very modest and bears little comparison to experimental results~\cite{smith2022electrically}. Moreover, the finite bias peaks melt rapidly with increasing temperature (left inset, fig~\ref{fig:finT}), leading to a broad featureless conductance. Such "spectral weight" transfer over large scales is indeed a characteristic of the weak/intermediate coupling regime. Hence, the finite bias conductance peak in the weak/intermediate coupling regime fails to explain the experimental results. As we will show below, the latter is best understood within the IPA framework from a strong correlation  perspective.

  \begin{figure}[tb]
       \centering
       \includegraphics[width=1.0\linewidth]{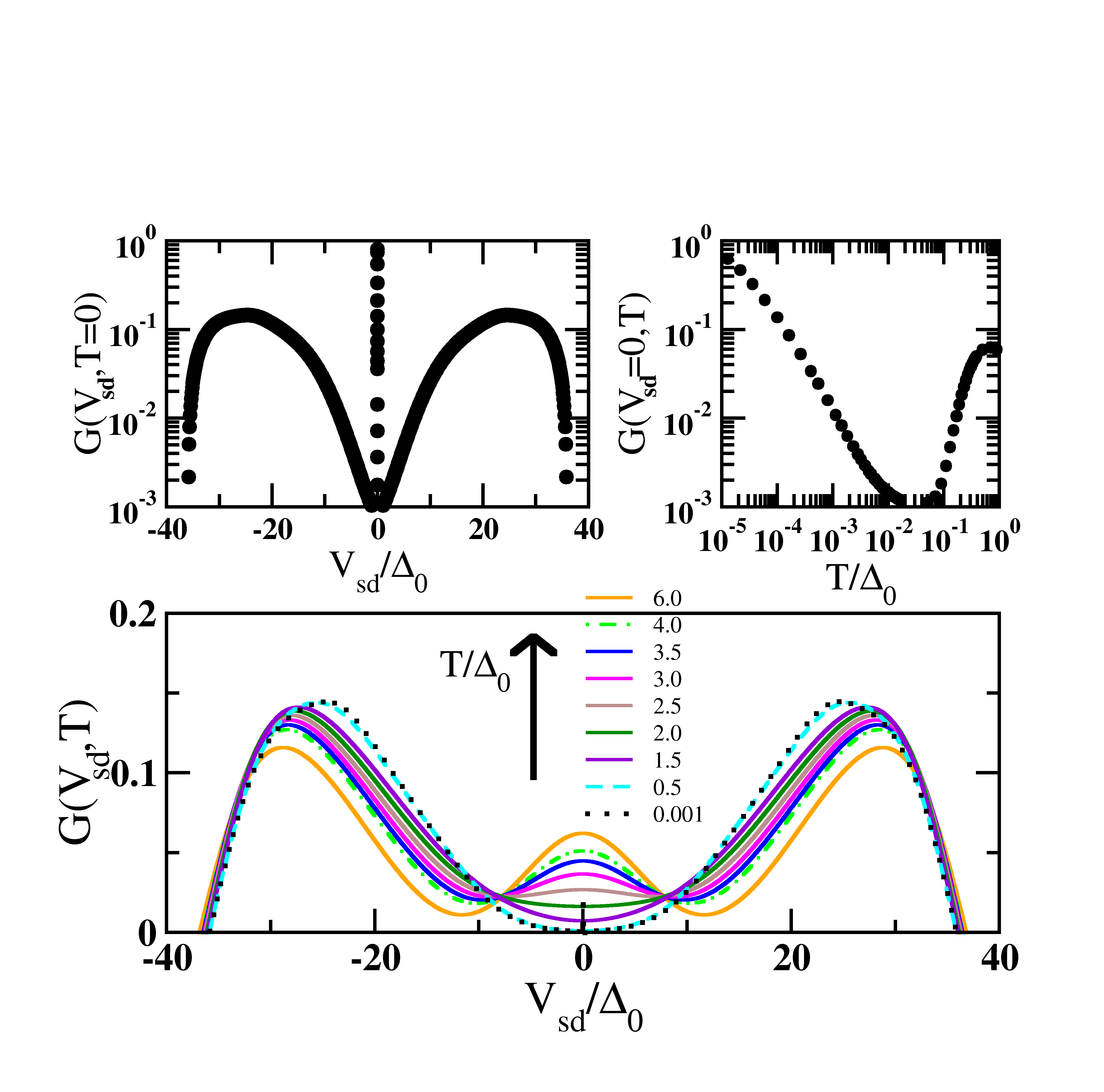}
       \caption{Scaled conductance, $G(V_{sd},T)$ as a function of bias, $V_{sd}/\Delta_0$ for various temperatures, $T/\Delta_0$ indicated as legends in the main panel. The parameters (except for bias) are the same as those considered in Fig.~\ref{fig:schem}(c).
       The left inset shows the zero temperature conductance as a function of scaled bias, while the right inset shows the ZBC, $G(V_{sd}=0, T)$ as a function
       of scaled temperature. }
       \label{fig:dbza}
   \end{figure}

 \subsubsection{Strong correlation regime and comparison to experiments} 
 As we noted earlier, the experimental results of Smith \etal\cite{smith2022electrically} are in clear disagreement with our results from the low/intermediate correlation regime. Thus, in this subsection, we will study the system in the strong correlation regime.
 The temperature interval over which the resistivity shows universal behavior when viewed on an absolute scale, i.e as $T/\Delta_0$ (and not as $T/T_K$) shrinks drastically in the strong interaction and strong spin-orbit coupling regime, because $T/\Delta_0 = (T/T_K)(T_K/\Delta_0)$, and the latter term is exponentially small in the strong coupling regime. In such a regime, a clear separation of scales happens in the conductivity as well as in the spectra. We show this in the top left panel of Fig.~\ref{fig:dbza}, where the conductance is shown as a function of bias, computed within IPA at $T=0$ for $U=20\Delta_0, \lambda=18\Delta_0$, using the Gaussian form of hybridization. 

The quasiparticle weight that determines the linear response regime, and the $U$ scale that determines the finite bias conductance peak show up as distinct features in the form of an extremely narrow zero bias peak (width $\sim Z_0\Delta_0\sim 10^{-3}\Delta_0$), and broad, finite bias peaks. The top right panel shows that the ZBC decreases rapidly with increasing temperature, and reaches a value three orders of magnitude smaller than the $T=0$ unitary limit at $T\sim 10^{-2}\Delta_0$. The ZBC follows the evolution of the dot spectral function and as  Fig.~\ref{fig:schem} shows, the melting of the Kondo peak manifests in the rapid decrease of the ZBC. However, as the temperature reaches non-universal values beyond $0.5\Delta_0$, the Hubbard bands become accessible, and the ZBC starts to rise again. The bottom panel shows the conductance as a function of bias for various temperatures in this rising ZBC regime. The two finite bias peaks are the only distinct features of the conductance in a wide temperature interval of  $T \in (10^{-3}\Delta_0, 0.5\Delta_0)$.
For $T\gtrsim 0.5\Delta_0$, consistent with the top right panel, the ZBC starts rising, filling in the gap between the finite bias peaks, and concomitantly the latter slightly diminishes in magnitude. An isosbestic point, commonly seen in many correlated systems, is visible around $V_{sd}\simeq 10\Delta_0$. In contrast to the results of the weak/intermediate coupling regime of the previous section, the finite bias peaks retain their form even at $T/\Delta_0=6.0$ as the bottom panel of Fig.~\ref{fig:finT} shows, and this behavior is characteristic of the strong coupling regime.

In a recent experiment by Smith \etal\cite{smith2022electrically}, the conductance through a point contact in an InGaAs/InAlAs heterostructure and a split-gate geometry was measured as a function of source-drain bias with varying in-plane and transverse magnetic fields. The highlight of this study was the tunability of the Kondo effect through the tuning of the Rashba spin-orbit coupling of the leads. The split-gate voltage, $V_{sg}$, was used to control the electron density, which in turn, determines the strength of the RSOC ($\lambda$). At a fixed temperature of about $25 {\rm mK}$, a single zero bias peak was observed at low $\lambda$, while a 'double zero bias peak' was observed at high $\lambda$ (see fig1(i) of Smith \etal\cite{smith2022electrically}). 

The zero bias peak is known to be related to the equilibrium Kondo scale, and in the absence of a magnetic field, the zero bias peak must be present in the conductance at $T=0$, no matter how strong the correlation is. But since the experiment is performed at a fixed and finite absolute temperature, the scenario becomes drastically different. As $\lambda$ is increased, the $T_K$ value decreases, so the ratio of $T/T_K$ increases exponentially. This implies that the
system with low $\lambda$ experiences a low $T/T_K$ ratio, while the system with high $\lambda$ is at a very high $T/T_K$ ratio for a given $T$. We have shown in Fig.~\ref{fig:dbza} that the central zero bias peak melts rapidly on a scale of $T_K$ at high $\lambda$, while the finite bias peaks retain their form. Thus, we conjecture that the experimental finding of a crossover from a single peak to a two-peak structure by increasing $\lambda$ is arising through a combined crossover of a weak to strong coupling (exponentially decreasing $T_K$) and from a low $T/T_K$ ratio to a high value of $T/T_K$ (due to a fixed $T$).

     \begin{figure}
       \centering
       \includegraphics[width=1.0\linewidth,trim=0 0 0 0, clip]{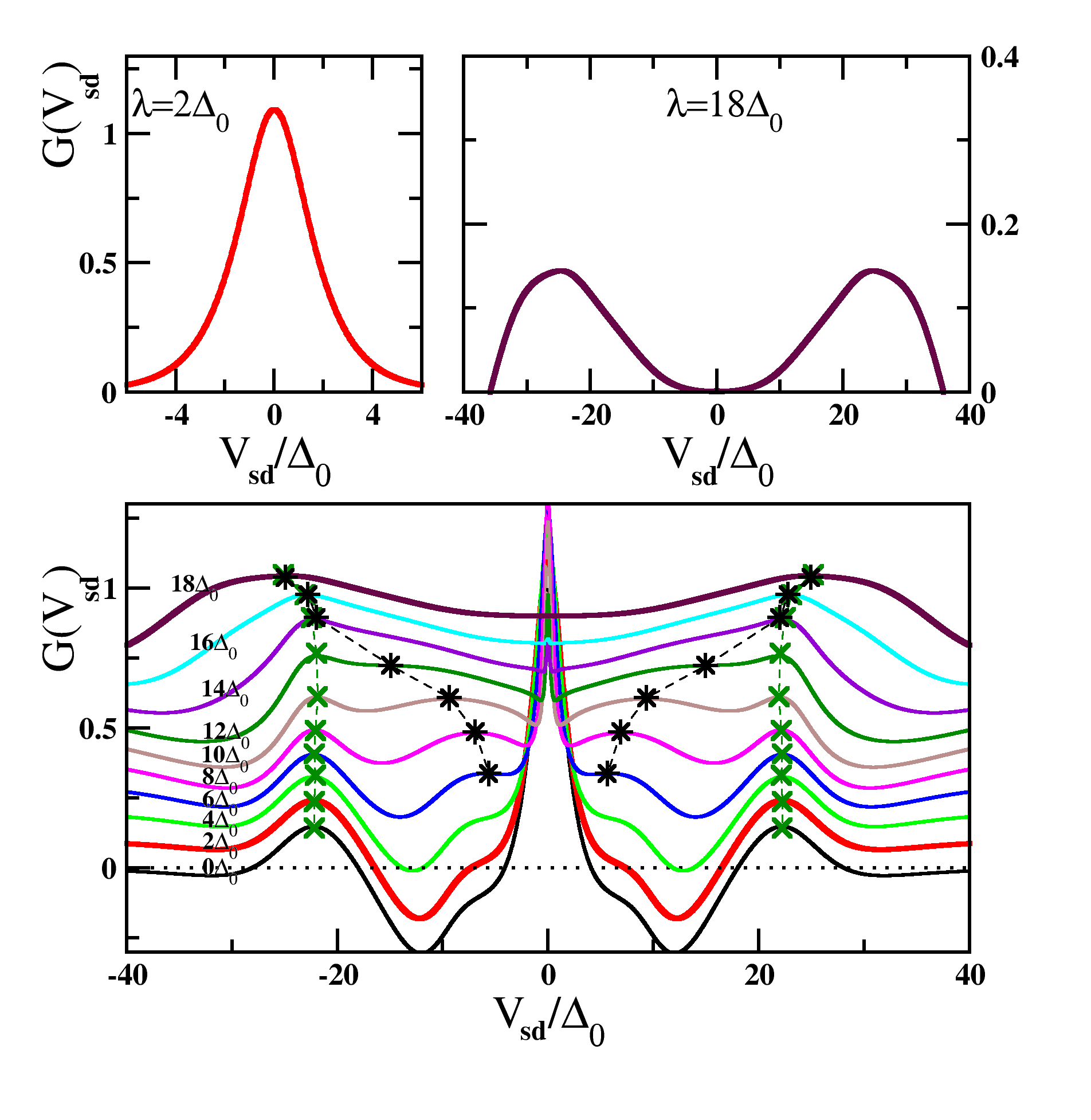}
       \caption{The conductance plotted as a function of $V_{sd}/\Delta_0$ for various values of SOC strengths as marked on the curves. The interaction and temperature are kept fixed at $U=20\Delta_0, T=0.05\Delta_0$ respectively. The crosses track the location of the $V_{sd}=\pm U$ peak, while the stars track a satellite feature of the zero bias peak. The left and right insets correspond to the low ($\lambda=2\Delta_0$) and high ($\lambda=18\Delta_0$) SOC strengths respectively. The insets show the different single and double peak conductance in these regimes.}
       \label{fig:fig1fg}
   \end{figure}
  
We consolidate this speculation in Fig.~\ref{fig:fig1fg}, where the top two panels show the conductance as a function of bias at low $\lambda=2\Delta_0$ (left panel) and high $\lambda=18\Delta_0$ (right panel) computed within IPA at a fixed temperature of $T=0.05\Delta_0$, and $U=20\Delta_0$ for a Gaussian hybridization. We see that a single zero bias peak has transformed into a double peak structure. We emphasize that the single zero bias peak is a universal feature implying that its form and temperature dependence are determined by the Kondo scale, but the finite bias peaks are non-universal features whose position, and form are determined by non-universal parameters such as $V_{sd},U$ and $\lambda$. The bottom panel shows the conductance for the same parameters as the top two panels, but with varying $\lambda$ marked as numbers. The conductance curves are shifted vertically by a constant number to provide clarity. The graph shows the gradual evolution of the single peak structure into a two-peak structure. The dashed line marks the evolution of these finite bias peaks (marked by crosses). Interestingly, the central peak is flanked by two satellite structures, which we identify visually (marked by stars). These peaks drift to higher $V_{sd}$ with increasing $\lambda$, and merge with the `Hubbard peaks', and continue the blue shift for higher $\lambda$ values. Similar features are observed in the experimental results (see Fig. 4(c) of Smith \etal~\cite{smith2022electrically}). Naturally, 
experimental results are far richer than what our simple model and the perturbative calculation reveal. However, the theoretical model used for explaining the experimental results~\cite{smith2022electrically} is exactly the same as the one used in the present work, and our results based on IPA may be viewed as complementary to those found by the quantum master equation approach. 
   
\section{Discussion and conclusions}  
We have investigated the interplay of electron-electron interactions, bias, and spin-orbit coupling on the conductance through an Anderson impurity, using a second-order Keldysh approach. We validated the method and our implementation through extensive benchmarks. The linear response regime and its relation with the equilibrium quasiparticle weight ($Z_0$) is explored, and the decrease of $Z_0$ with increasing spin-orbit coupling ($\lambda$),
for finite bandwidth hybridization functions, is highlighted and discussed in detail. This decrease turns out to be crucial for explaining the experimental observation of a crossover of a single zero bias peak into a double peak structure with increasing $\lambda$. The quantum master equation approach~\cite{smith2022electrically}, which was used for explaining the experimental results~\cite{smith2022electrically}, while recovering the split peak structure, addresses the finite field effects. However, the dependence of the Kondo scale on $\lambda$ was not incorporated, and finite temperature effects have not been addressed, and these two aspects have been incorporated and investigated in detail in our work.

One of the inferences made from the experimental results~\cite{smith2022electrically} is that the Kondo scale increases with increasing $\lambda$. This inference directly contradicts our results, since in our study the Kondo scale is represented by $Z_0\Delta_0$ which has been found to decrease with increasing $\lambda$ (Fig.~\ref{fig:scaled_int} and hence deserves a discussion. We note that the $T_K$ is determined experimentally for the experimental system at high $\lambda$ by increasing the magnetic field (parallel to the SOC field), which induces Zeeman splitting of the spin-degenerate levels, until the double peak form of the conductance converges to a single zero bias peak structure, and then fitting the temperature dependence of the ZBC. It is well known from equilibrium studies that the quasiparticle weight increases monotonically with increasing magnetic field\cite{logan2001magnetic}, so the determination of $T_K$ for the high magnetic field system is probably not representative of the $T_K$ for the zero-field system with the double peak conductance. In fact, for this system, if our results are any indication, then the $T_K$ is probably around the lower limit of the temperatures considered in the experiments. Increasing the magnetic field is akin to a crossover from strong to weak coupling regime, and hence is equivalent to decreasing spin-orbit coupling. And hence the merging of the double peak conductance to a single peak with increasing magnetic field may be viewed as decreasing $\lambda$ in Fig.~\ref{fig:dbza}.

 A 0.7 anomaly has been observed in the experimental results and we provide a brief discussion of this feature vis-a-vis our results. The 0.7 anomaly is seen as a plateau or shoulder in the conductivity when the gate voltage is varied. The origin of the 0.7 anomaly has been under intense study and has been attributed to the Kondo effect\cite{meir2002kondo} and more recently to van Hove ridges \cite{bauer2013,schimmel2017spin}. While the Kondo and van Hove ridge physics apply to quantum dots and 1D nanowire models, it has been shown that the low energy physics of the two systems are identical \cite{bauer2013}. In the present work, changing the gate voltage amounts to moving away from the p-h symmetric limit. The scope of the present paper is limited to the p-h symmetric limit, and broad featureless hybridization functions (hence ruling out van-Hove singularities), hence even in principle, the present work does not have any bearing on the physics of the 0.7 anomaly. 
 
While the IPA has been shown to work well in the parameter regime under consideration, it fails to capture many subtler aspects such as the exponential decrease of $Z_0$ with increasing interaction strength, which requires a method capable of capturing true strong coupling physics where spin fluctuations are incorporated non-perturbatively. The IPA has been generalized for finite magnetic fields~\cite{aligiamain} as well as for the p-h asymmetric case~\cite{aligiamain}, incorporating which will yield fresh insights into magnetotransport measurements and the influence of valence fluctuations, and possibly the 0.7 zero bias anomaly as well. We plan to pursue these directions.

\begin{acknowledgments}
The authors would like to thank Vinayak M.\ Kulkarni and Arijit Dutta for discussions.
We also acknowledge funding from JNCASR and the national supercomputing mission (DST/NSM/R\&D\_HPC\_Applications/2021/26).
\end{acknowledgments}

\appendix* 
\section{Derivation of an analytical expression for the quasiparticle weight}

    In order to get a better insight into the dependence of the self-energy on interactions, bias and SOC, we calculate an analytical expression for the quasiparticle weight using the second-order self-energy expression \eqref{eq:secnd_ordr_sigma_r}.
\begin{equation}
    \begin{split}
    \Sigma^r(\omega) &= U^2 D_0^3\int d\epsilon_1d\epsilon_2d\epsilon_3 \left(\underbrace{\frac{\tilde{f}(-\epsilon_1)\tilde{f}(-\epsilon_2)\tilde{f}(\epsilon_3)}{\omega^++\epsilon_3-\epsilon_2-\epsilon_1}}_{\Sigma^r_1}\right.\\
    &\left.+\underbrace{\frac{\tilde{f}(\epsilon_1)\tilde{f}(\epsilon_2)\tilde{f}(-\epsilon_3)}{\omega^++\epsilon_3-\epsilon_2-\epsilon_1}}_{\Sigma^r_2}\right)\,,
    \end{split}
\end{equation}
 where $D_0$ is the Hartree-corrected density of states at $\omega=0$. 
The two terms $\Sigma^r_{1,2}$ are not independent and we can show that $\Sigma^r_{2}(\omega)=(\Sigma^r_1(-\omega))^*$. When we consider the zero temperature limit and define $\rho_\Sigma=-1/\pi\Im{\Sigma^r_1(\omega)}$, the expression then
reduces to a product of Heaviside step functions given by
\begin{equation}
 \begin{split}
 \rho^r_{\Sigma1}(\omega) &= \frac{U^2 D^3_0}{8}\int d\epsilon_1d\epsilon_2 \left( \left[\underbrace{\Theta\left(V_{sd}/2-\epsilon_1 \right)}_{a_1}+ \underbrace{\Theta\left(-V_{sd}/2-\epsilon_1\right)}_{a_2}\right] \right. \\
 &\left.\left[\underbrace{\Theta\left(V_{sd}/2-\epsilon_2\right)}_{b_1}+\underbrace{\Theta\left(-V_{sd}/2-\epsilon_2\right)}_{b_2}\right]\right. \\
 &\left.\left[\underbrace{\Theta\left(V_{sd}/2+\omega-\epsilon_1-\epsilon_2\right)}_{c_1}+\underbrace{\Theta\left(-V_{sd}/2+\omega-\epsilon_1-\epsilon_2\right)}_{c_2}\right] \right)
 \end{split}
\end{equation}
The various products of the terms $a_i,b_i,c_i$ are tabulated below
\begin{equation}
\begin{array}{|c|c|}
     \hline
     {a_ib_ic_i} &  8\rho^r_{\Sigma1}|_{a_ib_ic_i}/(U^2D_0^3) \\
     \hline\hline
     {a_1b_1c_1}  & \omega^2/2+3\omega V_{sd}/2+V_{sd}^2+V_{sd}^2/8 \\
     {a_1b_1c_2}  & \omega^2/2+\omega V_{sd}/2+V_{sd}^2/8 \\
     {a_1b_2c_1}  & \omega^2/2+\omega V_{sd}/2+V_{sd}^2/8 \\
     {a_1b_2c_2}  & (\omega^2/2-\omega V_{sd}/2+V_{sd}^2/8)\Theta(\omega-V_{sd}/2) \\
     {a_2b_1c_1}  & \omega^2/2+\omega V_{sd}/2+V_{sd}^2/8 \\
     {a_2b_1c_2}  & (\omega^2/2-\omega V_{sd}/2+V_{sd}^2/8)\Theta(\omega-V_{sd}/2) \\
     {a_2b_2c_1}  & (\omega^2/2-\omega V_{sd}/2+V_{sd}^2/8)\Theta(\omega-V_{sd}/2) \\
     {a_2b_2c_2}  & (\omega^2/2-3\omega V_{sd}/2+V_{sd}^2+V_{sd}^2/8)\Theta(\omega-3V_{sd}/2) \\ 
     \hline
\end{array}
\end{equation}
Adding all the terms we get 
\begin{equation}
    \rho^r_{\Sigma} = \frac{U^2 D^3_0}{2}\left[\omega^2+\frac{3V_{sd}^2}{4}\right]\,.
    \label{eqn:rhoself}
\end{equation}
We can use the Kramer-Kr\"{o}nig transform to calculate the real part of the self-energy as
\begin{equation}
    \Re\Sigma^r(\omega) = \frac{1}{\pi}\mathcal{P}\int d\omega'\frac{\Im\Sigma^r(\omega')}{\omega'-\omega}\,
\end{equation}
which we can write by considering a high energy cutoff $\Lambda$ as
\begin{equation}
    \Re\Sigma^r(\omega) = \frac{U^2 D_0^3}{2}\int_{-\Lambda}^\Lambda d\omega' \frac{\omega'^2+3V_{sd}^2/4}{\omega'-\omega}\,.
\end{equation}
Considering particle-hole symmetry and truncating terms of $\mathcal{O}(\omega^2)$, we get the expression
\begin{equation}
    \Re\Sigma^r(\omega) = -\frac{U^2 D^3_0}{\Lambda}\left[\Lambda^2-\frac{3V_{sd}^2}{4}\right]\omega\,.
\end{equation}
Using the definition of the quasiparticle weight as $\Re\Sigma^r(\omega) = \omega(1-\frac{1}{Z}) $ we get the analytical expression (Eq. \eqref{eqn:zmain} in the main text) as
\begin{equation}
    Z = \left(1+\frac{U^2 D^3_0}{\Lambda}\left[\Lambda^2-\frac{3V_{sd}^2}{4}\right]\right)^{-1}\,.
    \label{eqn:analyticalz}
\end{equation}
 
\newpage
\bibliographystyle{apsrev4-1}
\bibliography{ipaSOC}
\end{document}